\documentclass[12pt]{article}
\usepackage{epsfig}
\begin{document}
\title{Pion-induced production of $\Lambda(1520)$ hyperons on nuclei near threshold}
\author{E. Ya. Paryev$^{1,2}$, Yu. T. Kiselev$^2$\\
{\it $^1$Institute for Nuclear Research, Russian Academy of Sciences,}\\
{\it Moscow 117312, Russia}\\
{\it $^2$Institute for Theoretical and Experimental Physics,}\\
{\it Moscow 117218, Russia}}

\renewcommand{\today}{}
\maketitle

\begin{abstract}
We study the pion-induced inclusive $\Lambda(1520)$ hyperon production from nuclei near threshold
within a nuclear spectral function approach accounting for incoherent primary $\pi^-$ meson--proton
${\pi^-}p \to K^0\Lambda(1520)$ production processes. We calculate the absolute differential and total cross
sections for the production of $\Lambda(1520)$ hyperons off carbon and tungsten nuclei at laboratory angles of
0$^{\circ}$--10$^{\circ}$, 10$^{\circ}$--45$^{\circ}$ and 45$^{\circ}$--85$^{\circ}$ by $\pi^-$ mesons
with momentum of 1.7 GeV/c as well as their relative (transparency ratio) differential and integral
yields from these nuclei within four scenarios for their total in-medium width. We demonstrate that
these absolute observables, contrary to the relative ones, reveal some sensitivity to
the $\Lambda(1520)$ in-medium width. Therefore, their measurement in a dedicated experiment at
the GSI pion beam facility will allow to shed light on this width.
\end{abstract}

\newpage

\section*{1 Introduction}

\hspace{0.5cm} The study of the renormalization of the properties, masses and widths, of 
light vector mesons $\rho$, $\omega$, $\phi$, kaons and antikaons, $\eta$, $\eta^{\prime}$ and $J/\psi$
mesons in nuclei has received considerable interest in recent years (see, for example, [1--9]).
The in-medium properties of hyperons at finite density have also become a topic of intense current theoretical
investigations [10--16]. Thus, for instance, the medium modifications of $\Sigma(1385)$ and $\Lambda(1520)$
hyperons have been studied within chiral unitary hadronic theory [10, 11]. Very small mass shifts of about
-40 and -20 MeV have been predicted for the $\Sigma(1385)$ and $\Lambda(1520)$ hyperons, respectively, at rest
at normal nuclear matter density $\rho_0$. Moreover, a moderate increase of the width of
$\Sigma(1385)$ by the factor $\sim$ 2--3 at density $\rho_0$ compared to the free width $\sim$ 35 MeV and
appreciable increase of the width of $\Lambda(1520)$ hyperons at this density by the factor $\sim$ 5 with
respect to the free width of 15.6 MeV has been calculated. The influence of the $\Lambda(1520)$ hyperon 
in-medium width on its
yield from $pA$ and ${\gamma}A$ reactions has been analyzed within collisional models [17--19] using an
eikonal approximation. It has been shown that both the momentum dependence of the absolute $\Lambda(1520)$
hyperon yield and the $A$ dependence of its relative yield are quite sensitive to the $\Lambda(1520)$
in-medium width.

 Valuable information on the in-medium properties of $\Lambda(1520)$ hyperons, complementary to that from
proton--nucleus and photon--nucleus collisions, can be inferred from pion--nucleus reactions.
In this context, the main aim of the present study is to give the predictions
for the absolute differential and total cross sections for production of $\Lambda(1520)$ hyperons in
${\pi^-}^{12}C \to {\Lambda(1520)}X$ and ${\pi^-}^{184}W \to {\Lambda(1520)}X$ reactions at 1.7 GeV/c
incident pion momentum as well as for their relative (transparency ratio) differential and integral yields
from these reactions within different scenarios for $\Lambda(1520)$ total in-medium width.
These nuclear targets and this initial beam momentum were adopted in recent measurements [20]
of $\pi^-$ meson-induced $\phi$ meson production at the GSI pion beam facility using the HADES
spectrometer and, therefore, can be employed in studying the ${\pi^-}A \to {\Lambda(1520)}X$ interactions here.
The calculations are based on a first-collision model, developed in [21] for the analysis of the inclusive
$\phi$ meson production data [20] and extended to account for different scenarios for the
$\Lambda(1520)$ in-medium width. These calculations can be used as an important tool for determining
the width from the data which could be taken in a dedicated experiment at the GSI pion beam facility.

\section*{2 Direct  $\Lambda(1520)$ hyperon production mechanism}

\hspace{0.5cm} The $\Lambda(1520)$ hyperons can be produced directly
in $\pi^-A$ ($A=^{12}$C and $^{184}$W) reactions at incident momentum of 1.7 GeV/c
in the following $\pi^-p$ elementary process with the lowest free production threshold momentum (1.68 GeV/c)
\footnote{$^)$ We can neglect the processes $\pi^-N \to {K}\Lambda(1520){\pi}$ with one pion in the final state
at the incident pion momentum of 1.7 GeV/c, because they are energetically suppressed due to the fact that
this momentum is less than their production threshold momentum of 1.98 GeV/c in vacuum $\pi^-N$ collisions.
Moreover, taking into account the results of the study [21] of pion-induced
$\phi$ meson production on $^{12}$C and $^{184}$W nuclei at beam momentum of 1.7 GeV/c,
we ignore in the present work by analogy with [21] the secondary pion--nucleon ${\pi}N \to {K}\Lambda(1520)$
production processes.}$^)$
\begin{equation}
\pi^-+p \to K^0+\Lambda(1520).
\end{equation}
Taking into consideration that the in-medium threshold energy
\footnote{$^)$ Determining the strength of the $\Lambda(1520)$ production cross sections in $\pi^-A$ collisions
near threshold (cf. [21]).}$^)$
of the process (1) looks like that for the
free final particles due to the cancelation of nuclear scalar potentials $U_{K^0}\approx+20$ MeV [5] and
$U_{\Lambda(1520)}\approx-20$ MeV [10], felt by the $K^0$ mesons and low-momentum $\Lambda(1520)$ hyperons
at saturation density $\rho_0$, as well as for the reason of reducing the possible uncertainty of our
calculations due to use in them of the model $\Lambda(1520)$ self-energy [10] at finite momenta studied
in the present work, we will ignore here the medium modification of the outgoing hadron masses at these momenta.

  Since the $\Lambda(1520)$--nucleon elastic cross section
is expected to be small similar to the ${\Lambda}N$ elastic cross section at these
momenta [22], we will neglect quasielastic ${\Lambda(1520)}N$ rescatterings in the present study.
Then, accounting for the absorption of the incident pion and $\Lambda(1520)$ hyperon in nuclear matter
as well as using the results given in [21], we represent the inclusive differential
cross section for the production of ${\Lambda(1520)}$ hyperons with vacuum momentum ${\bf p}_{\Lambda(1520)}$
(or ${\bf p}_{\Lambda^*}$) on nuclei in the direct process (1) as follows:
\begin{equation}
\frac{d\sigma_{{\pi^-}A \to {\Lambda(1520)}X}^{({\rm prim})}
({\bf p}_{\pi^-},{\bf p}_{\Lambda^*})}
{d{\bf p}_{{\Lambda^*}}}=I_{V}[A,\theta_{{\Lambda^*}}]
\left(\frac{Z}{A}\right)\left<\frac{d\sigma_{{\pi^-}p\to K^0{{\Lambda(1520)}}}({\bf p}_{\pi^-},
{\bf p}_{{\Lambda^*}})}{d{\bf p}_{{\Lambda^*}}}\right>_A,
\end{equation}
where
\begin{equation}
I_{V}[A,\theta_{{\Lambda^*}}]=A\int\limits_{0}^{R}r_{\bot}dr_{\bot}
\int\limits_{-\sqrt{R^2-r_{\bot}^2}}^{\sqrt{R^2-r_{\bot}^2}}dz
\rho(\sqrt{r_{\bot}^2+z^2})
\exp{\left[-\sigma_{{\pi^-}N}^{\rm tot}A\int\limits_{-\sqrt{R^2-r_{\bot}^2}}^{z}
\rho(\sqrt{r_{\bot}^2+x^2})dx\right]}
\end{equation}
$$
\times
\int\limits_{0}^{2\pi}d{\varphi}\exp{\left[-
\int\limits_{0}^{l(\theta_{{\Lambda^*}},\varphi)}\frac{dx}
{\lambda_{\Lambda^*}(\sqrt{x^2+2a(\theta_{{\Lambda^*}},\varphi)x+b+R^2})}\right]},
$$
\begin{equation}
a(\theta_{{\Lambda^*}},\varphi)=z\cos{\theta_{{\Lambda^*}}}+
r_{\bot}\sin{\theta_{{\Lambda^*}}}\cos{\varphi},\\\
b=r_{\bot}^2+z^2-R^2,
\end{equation}
\begin{equation}
l(\theta_{{\Lambda^*}},\varphi)=\sqrt{a^2(\theta_{{\Lambda^*}},\varphi)-b}-
a(\theta_{{\Lambda^*}},\varphi),
\end{equation}
\begin{equation}
\lambda_{\Lambda^*}(|{\bf r}|)=\frac{p_{\Lambda^*}}{M_{\Lambda^*}\Gamma_{\Lambda^*}(|{\bf r}|)}
\end{equation}
and
\begin{equation}
\left<\frac{d\sigma_{{\pi^-}p\to K^0{\Lambda(1520)}}({\bf p}_{\pi^-},{\bf p}_{\Lambda^*})}
{d{\bf p}_{\Lambda^*}}\right>_A=
\int\int
P_A({\bf p}_t,E)d{\bf p}_tdE
\end{equation}
$$
\times
\left\{\frac{d\sigma_{{\pi^-}p\to K^0{\Lambda(1520)}}[\sqrt{s},M_{\Lambda^*},
m_{K^0},{\bf p}_{\Lambda^*}]}
{d{\bf p}_{\Lambda^*}}\right\},
$$
\begin{equation}
  s=(E_{\pi^-}+E_t)^2-({\bf p}_{\pi^-}+{\bf p}_t)^2,
\end{equation}
\begin{equation}
   E_t=M_A-\sqrt{(-{\bf p}_t)^2+(M_{A}-m_{N}+E)^{2}}.
\end{equation}
Here,
$d\sigma_{{\pi^-}p\to K^0{\Lambda(1520)}}[\sqrt{s},M_{{\Lambda^*}},m_{K^0},{\bf p}_{\Lambda^*}]
/d{\bf p}_{\Lambda^*}$
is the off-shell inclusive differential cross section for the production of ${\Lambda(1520)}$ hyperon and
$K^0$ meson with free masses $M_{{\Lambda^*}}$ and $m_{K^0}$, respectively. 
The $\Lambda(1520)$ hyperon is produced
with vacuum momentum ${\bf p}_{{\Lambda^*}}$ in reaction (1) at the ${\pi^-}p$ center-of-mass energy $\sqrt{s}$.
$E_{\pi^-}$ and ${\bf p}_{\pi^-}$ are the total energy and momentum of the incident pion
($E_{\pi^-}=\sqrt{m^2_{\pi}+{\bf p}^2_{\pi^-}}$, $m_{\pi}$ is the free space pion mass);
$\rho({\bf r})$ and $P_A({\bf p}_t,E)$ are the local nucleon density and the
spectral function of the target nucleus A normalized to unity;
${\bf p}_t$ and $E$ are the internal momentum and removal energy of the struck target proton
involved in the collision process (1); $\sigma_{{\pi^-}N}^{\rm tot}$ is the total cross section of the
free ${\pi^-}N$ interaction (we use in our calculations the value of $\sigma_{{\pi^-}N}^{\rm tot}=35$ mb
for initial pion momentum of 1.7 GeV/c [21]); $\Gamma_{\Lambda^*}(|{\bf r}|)$ is the total ${\Lambda(1520)}$
width in its rest frame, taken at the point ${\bf r}$ inside the nucleus and at the pole mass $M_{\Lambda^*}$;
$Z$ and $A$ are the numbers of protons and nucleons in
the target nucleus, and $M_{A}$  and $R$ are its mass and radius; $m_N$ is the free space nucleon mass;
and $\theta_{\Lambda^*}$ is the polar angle of
vacuum momentum ${\bf p}_{{\Lambda^*}}$ in the laboratory system with z-axis directed along the momentum
${\bf p}_{{\pi^-}}$ of the incoming pion beam.

        For the nuclear density $\rho({\bf r})$ of the $^{12}$C and $^{184}$W
nuclei of interest, we have adopted, correspondingly, the harmonic oscillator
and the Woods-Saxon distributions given in Ref. [21].
For these target nuclei, the nuclear spectral function $P_A({\bf p}_t,E)$ was taken from Refs.~[21, 23--25].

   Following [21], we suppose that the off-shell differential cross section \\
$d\sigma_{{\pi^-}p\to K^0{\Lambda(1520)}}[\sqrt{s},M_{\Lambda^*},m_{K^0},{\bf p}_{\Lambda^*}]
/d{\bf p}_{\Lambda^*}$ for $\Lambda(1520)$ production in process (1) is equivalent to
the respective on-shell cross section calculated for the off-shell kinematics of this process
as well as for the final $\Lambda(1520)$ and kaon free space masses $M_{\Lambda^*}$ and $m_{K^0}$,
respectively. Accounting for Eq.~(14) from [21], we obtain the following expression for this cross section:
\begin{equation}
\frac{d\sigma_{{\pi^{-}}p \to K^0{\Lambda(1520)}}[\sqrt{s},M_{\Lambda^*},m_{K^0},
{\bf p}_{\Lambda^*}]}{d{\bf p}_{\Lambda^*}}=
\frac{\pi}{I_2[s,M_{\Lambda^*},m_{K^0}]E_{\Lambda^*}}
\end{equation}
$$
\times
\frac{d\sigma_{{\pi^{-}}p \to K^0{\Lambda(1520)}}(\sqrt{s},M_{\Lambda^*},m_{K^0},\theta^*_{\Lambda^*})}
{d{\bf \Omega}^*_{\Lambda^*}}
$$
$$
\times
\frac{1}{(\omega+E_t)}\delta\left[\omega+E_t-\sqrt{m_{K^0}^2+({\bf Q}+{\bf p}_t)^2}\right],
$$
where
\begin{equation}
I_2[s,M_{\Lambda^*},m_{K^0}]=\frac{\pi}{2}
\frac{\lambda[s,M_{\Lambda^*}^{2},m_{K^0}^{2}]}{s},
\end{equation}
\begin{equation}
\lambda(x,y,z)=\sqrt{{\left[x-({\sqrt{y}}+{\sqrt{z}})^2\right]}{\left[x-
({\sqrt{y}}-{\sqrt{z}})^2\right]}},
\end{equation}
\begin{equation}
\omega=E_{\pi^-}-E_{\Lambda^*}, \,\,\,\,{\bf Q}={\bf p}_{\pi^-}-{\bf p}_{\Lambda^*},\,\,\,\,
E_{\Lambda^*}=\sqrt{M_{\Lambda^*}^{2}+{\bf p}^2_{\Lambda^*}}.
\end{equation}
Here,
$d\sigma_{{\pi^{-}}p \to K^0{\Lambda(1520)}}(\sqrt{s},M_{\Lambda^*},m_{K^0},\theta^*_{\Lambda^*})
/d{\bf \Omega}^*_{\Lambda^*}$
is the off-shell differential cross section for the production of $\Lambda(1520)$ hyperons
in reaction (1) under the polar angle $\theta^*_{\Lambda^*}$ in the ${\pi^-}p$ c.m.s. This cross section
is assumed to be isotropic in our model calculations of $\Lambda(1520)$ hyperon production in pion--nucleus interactions:
\begin{equation}
\frac{d\sigma_{{\pi^{-}}p \to K^0{\Lambda(1520)}}(\sqrt{s},M_{\Lambda^*},m_{K^0},\theta^*_{\Lambda^*})}
{d{\bf \Omega}^*_{\Lambda^*}}=\frac{\sigma_{{\pi^{-}}p \to K^0{\Lambda(1520)}}(\sqrt{s},\sqrt{s_{\rm th}})}{4\pi}.
\end{equation}
Here, $\sigma_{{\pi^{-}}p \to K^0{\Lambda(1520)}}(\sqrt{s},\sqrt{s_{\rm th}})$ is the "in-medium" total cross section
of reaction (1) having the threshold energy $\sqrt{s_{\rm th}}=M_{\Lambda^*}+m_{K^0}=2.017$ GeV.
According to the aforesaid, it is equivalent to the vacuum cross section
$\sigma_{{\pi^{-}}p \to K^0{\Lambda(1520)}}(\sqrt{s},\sqrt{s_{\rm th}})$, in which the free collision energy
$s=(E_{\pi^-}+m_N)^2-{\bf p}_{\pi^-}^2$ is replaced by the in-medium expression (8).
For the vacuum total cross section
$\sigma_{{\pi^{-}}p \to K^0{\Lambda(1520)}}(\sqrt{s},\sqrt{s_{\rm th}})$ we have employed the
following parametrization suggested in Ref. [19]:
\begin{equation}
\sigma_{{\pi}^-p \to K^0{\Lambda(1520)}}(\sqrt{s},\sqrt{s_{\rm th}})=\left\{
\begin{array}{ll}
	123\left(\sqrt{s}-\sqrt{s_{\rm th}}\right)^{0.47}~[{\rm {\mu}b}]
	&\mbox{for $0 < \sqrt{s}-\sqrt{s_{\rm th}} \le 0.427~{\rm GeV}$}, \\
	&\\
                   26.6/\left(\sqrt{s}-\sqrt{s_{\rm th}}\right)^{1.33}~[{\rm {\mu}b}]
	&\mbox{for $\sqrt{s}-\sqrt{s_{\rm th}} > 0.427~{\rm GeV}$}.
\end{array}
\right.	
\end{equation}
\begin{figure}[htb]
\begin{center}
\includegraphics[width=12.0cm]{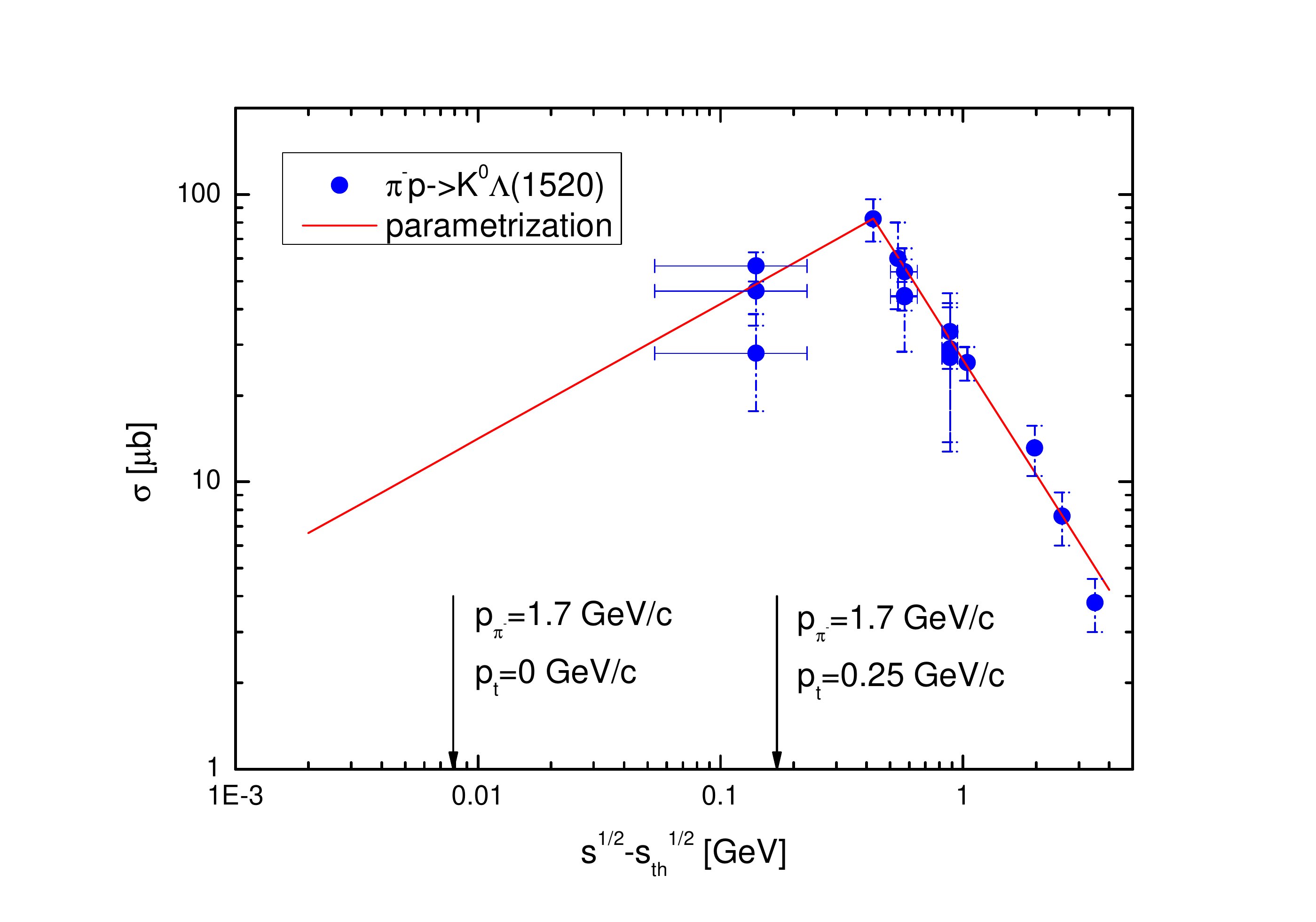}
\vspace*{-2mm} \caption{(color online) Total cross section for the reaction $\pi^-p \to K^0{\Lambda(1520)}$
as a function of the excess energy $\sqrt{s}-\sqrt{s_{\rm th}}$. The left and right arrows indicate
the excess energies $\sqrt{s}-\sqrt{s_{\rm th}}$=7.9 MeV and $\sqrt{s}-\sqrt{s_{\rm th}}$=171 MeV
corresponding to the incident pion momentum of $|{\bf p}_{\pi^-}|=1.7$ GeV/c and a
free target proton at rest and a target proton bound in $^{12}$C by 16 MeV
and having momentum of 250 MeV/c directed opposite to the incoming pion beam.
For the rest of notation see text.}
\label{void}
\end{center}
\end{figure}
As seen from Fig. 1, the parametrization (15) (solid line) fits well the existing set of experimental
data (full circles) [26] for the ${\pi^-}p \to K^0{\Lambda(1520)}$ reaction.
One can also see that the on-shell cross section $\sigma_{{\pi}^-p \to K^0{\Lambda(1520)}}$
amounts approximately to 13 $\mu$b for the initial pion momentum of 1.7 GeV/c and a free target proton
being at rest. The off-shell cross section $\sigma_{{\pi}^-p \to K^0{\Lambda(1520)}}$, calculated
in line with Eqs. (8), (9), (15) for a pion momentum of 1.7 GeV/c and a target proton bound in $^{12}$C 
by 16 MeV and having relevant internal momentum of 250 MeV/c, is about 53 $\mu$b. This
offers the possibility of measuring the ${\Lambda(1520)}$ yield in $\pi^-A$ reactions at the 
very near-threshold beam momentum of 1.7 GeV/c at the GSI pion beam facility with sizeable strength.
\begin{figure}[!h]
\begin{center}
\includegraphics[width=12.0cm]{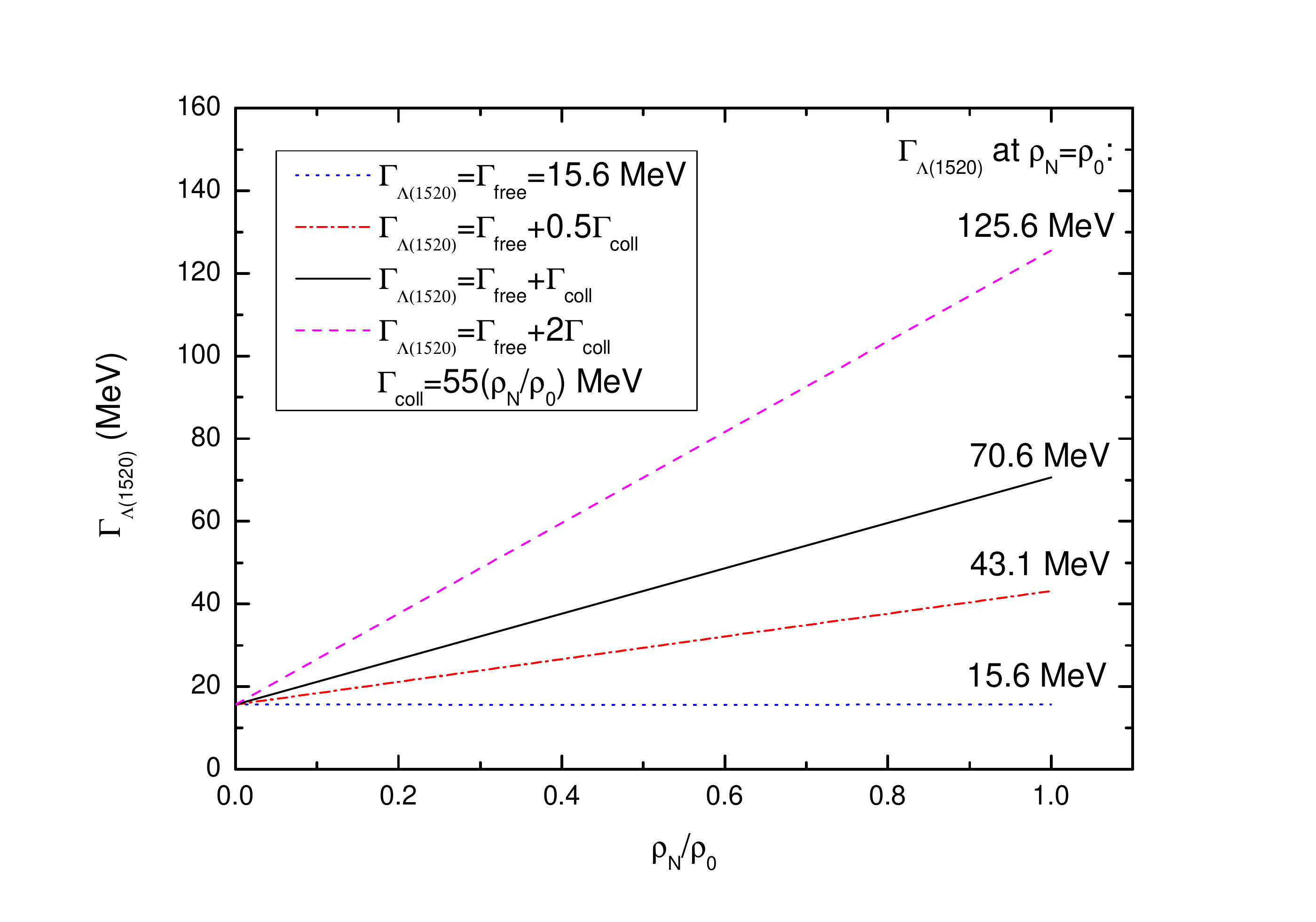}
\vspace*{-2mm} \caption{(color online) $\Lambda(1520)$ hyperon total width in its rest frame as a function
of the density. For notation see text.}
\label{void}
\end{center}
\end{figure}
\begin{figure}[!h]
\begin{center}
\includegraphics[width=16.0cm]{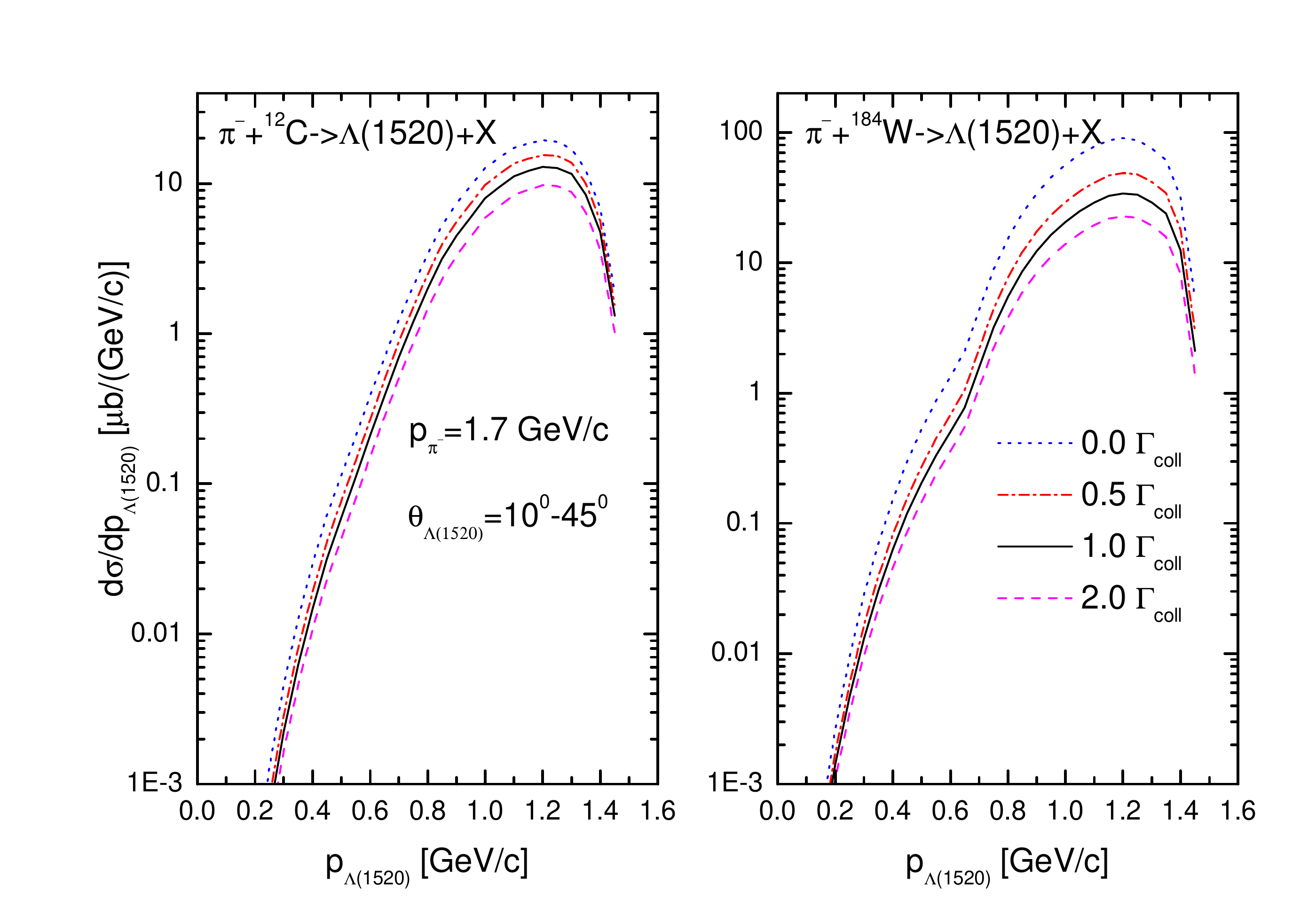}
\vspace*{-2mm} \caption{(color online) Momentum differential cross sections for the production of $\Lambda(1520)$
hyperons from the primary ${\pi^-}p \to K^0{\Lambda(1520)}$ channel in the laboratory polar angular range of
10$^{\circ}$--45$^{\circ}$ in the interaction of $\pi^-$ mesons of momentum of 1.7 GeV/c with $^{12}$C
(left) and $^{184}$W (right) nuclei, calculated within different scenarios for the total $\Lambda(1520)$
hyperon in-medium width in which its collisional width was multiplied by the factors indicated in the inset.}
\label{void}
\end{center}
\end{figure}
\begin{figure}[!h]
\begin{center}
\includegraphics[width=18.0cm]{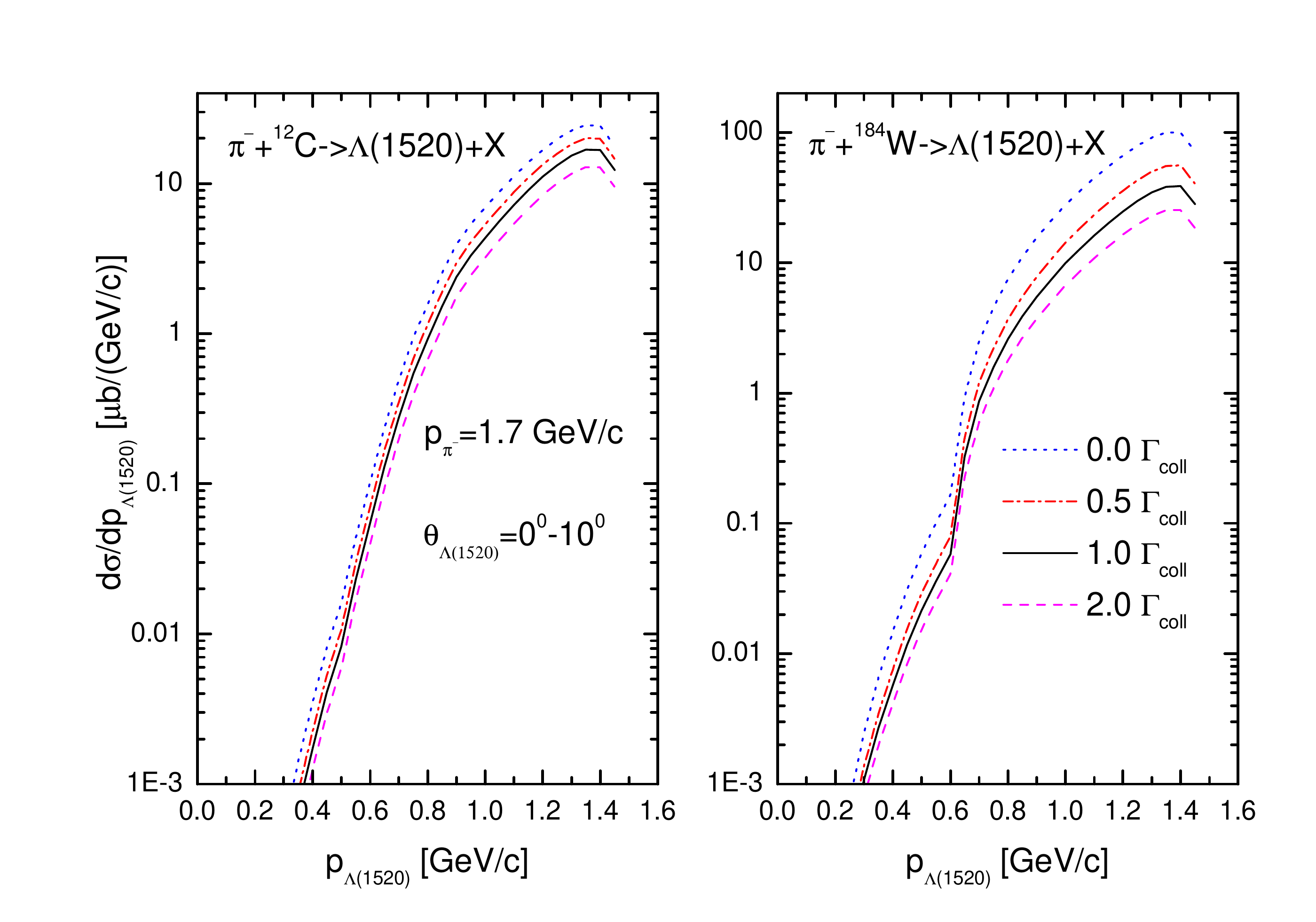}
\vspace*{-2mm} \caption{(color online) The same as in Fig.3 but for the laboratory polar angular range of
0$^{\circ}$--10$^{\circ}$.}
\label{void}
\end{center}
\end{figure}
\begin{figure}[!h]
\begin{center}
\includegraphics[width=18.0cm]{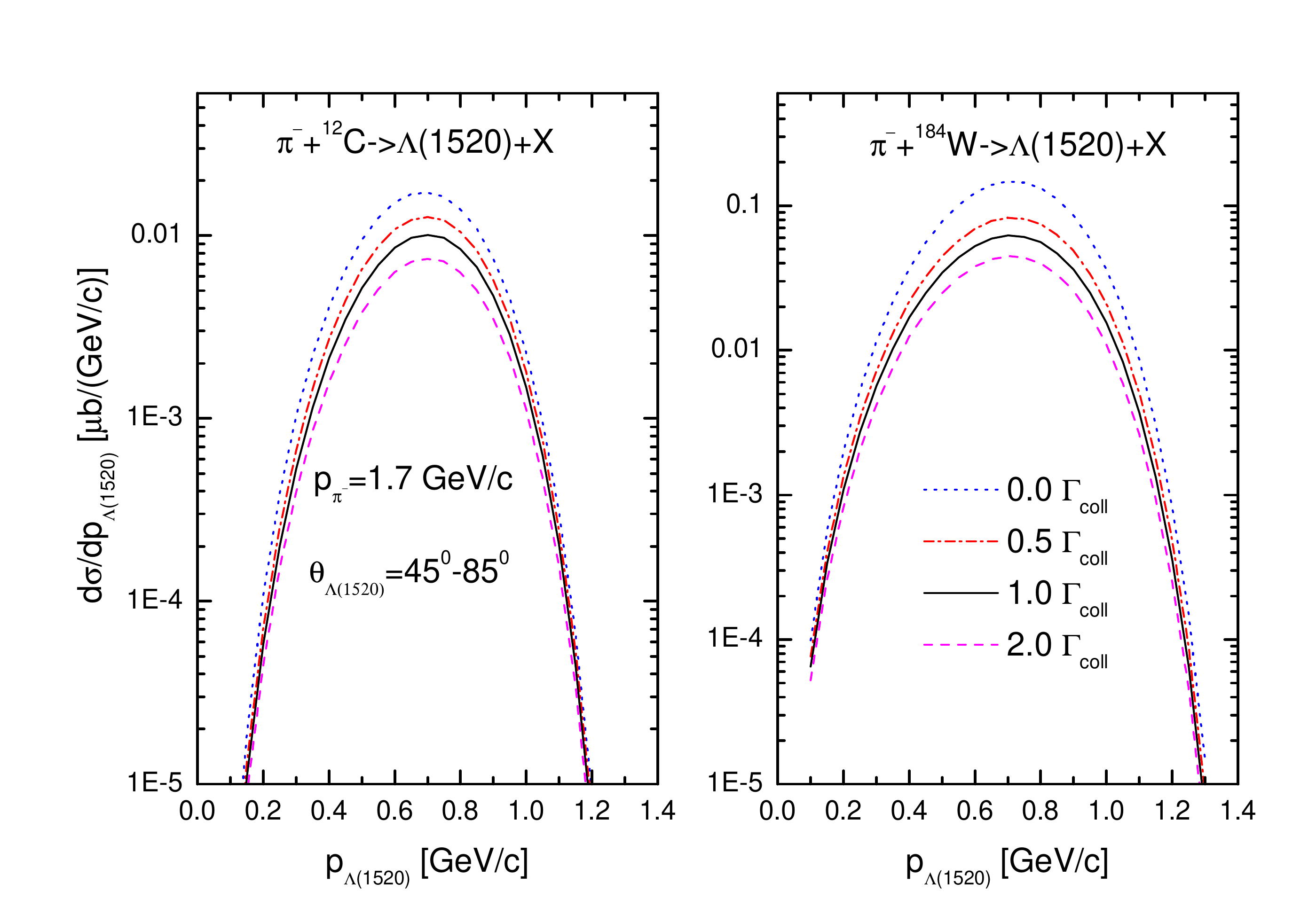}
\vspace*{-2mm} \caption{(color online) The same as in Fig.3 but for the laboratory polar angular range of
45$^{\circ}$--85$^{\circ}$.}
\label{void}
\end{center}
\end{figure}
\begin{figure}[!h]
\begin{center}
\includegraphics[width=18.0cm]{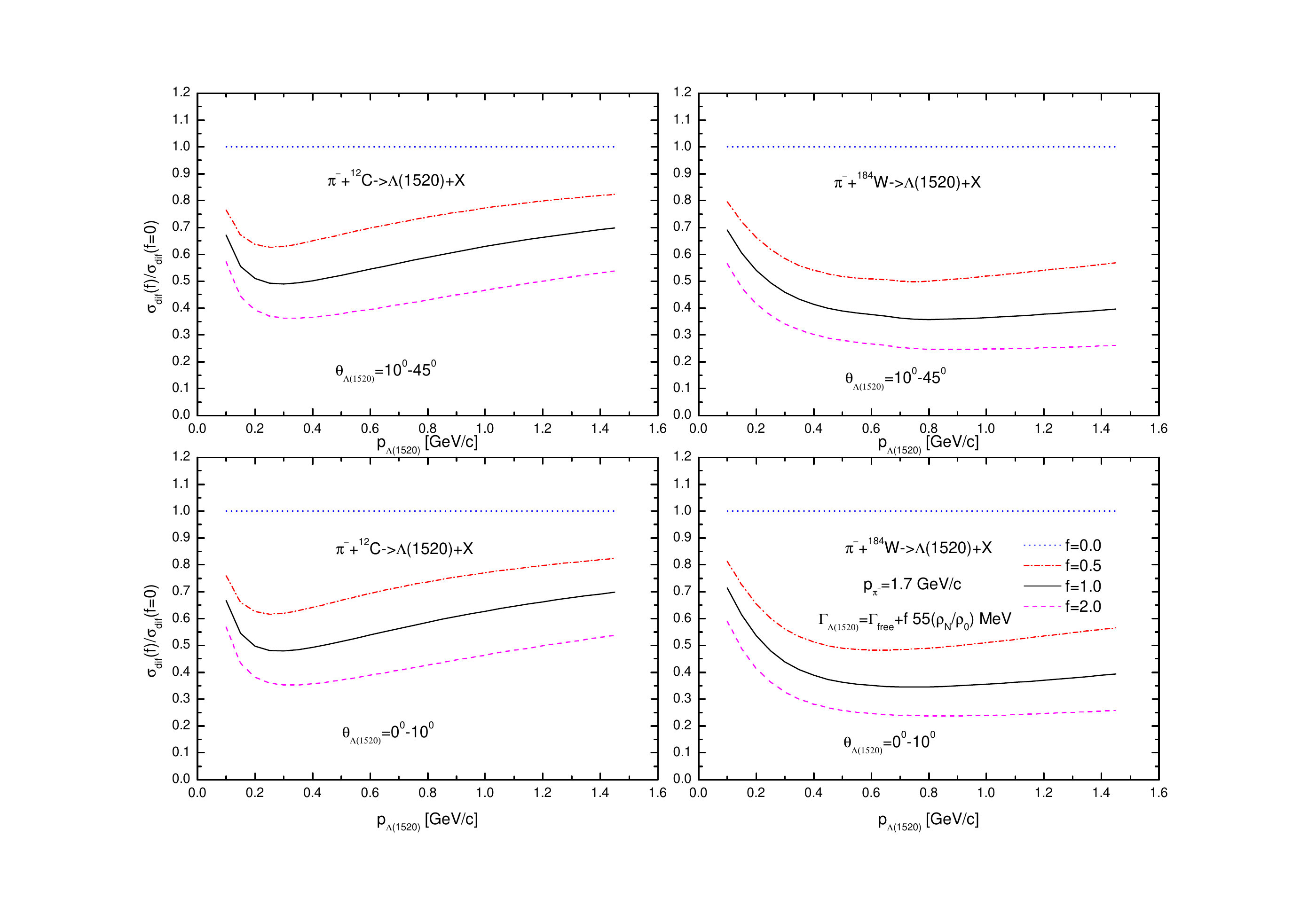}
\vspace*{-2mm} \caption{(color online) Ratio between the differential cross sections of $\Lambda(1520)$
production on $^{12}$C and $^{184}$W target nuclei at
laboratory angles of 10$^{\circ}$--45$^{\circ}$ (upper two panels) and 0$^{\circ}$--10$^{\circ}$
(lower two panels) by 1.7 GeV/c $\pi^-$ mesons in the primary ${\pi^-}p \to K^0{\Lambda(1520)}$ reactions
calculated within different scenarios for the total $\Lambda(1520)$
hyperon in-medium width in which its collisional width was multiplied by the factors indicated in the inset,
and the same cross sections, obtained in the scenario where the absorption of $\Lambda(1520)$ hyperons in
nuclear matter is governed by their free width (dotted curve in Fig. 2),
as a function of $\Lambda(1520)$ momentum.}
\label{void}
\end{center}
\end{figure}
\begin{figure}[!h]
\begin{center}
\includegraphics[width=18.0cm]{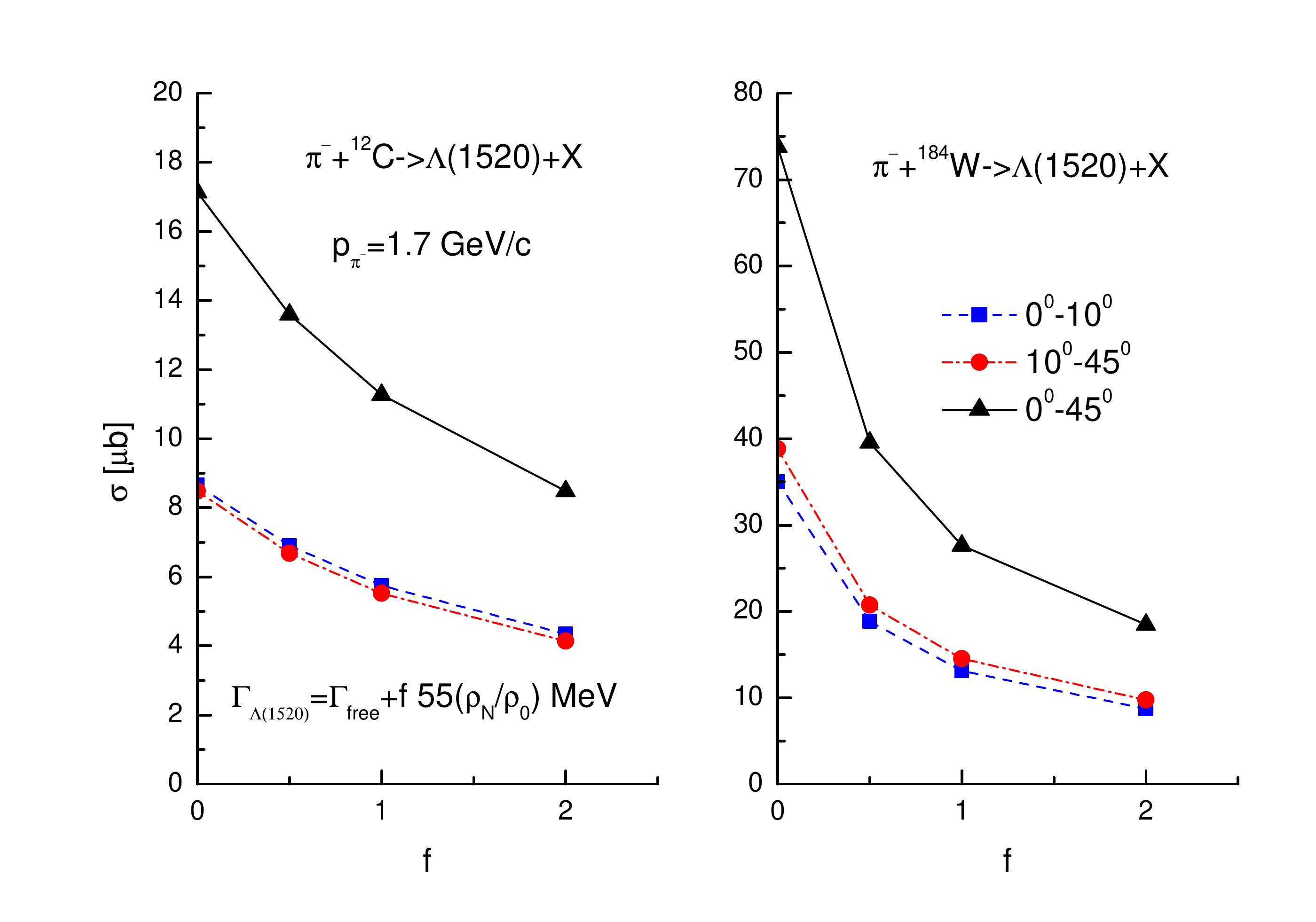}
\vspace*{-2mm} \caption{(color online) The total cross sections for the production of $\Lambda(1520)$
hyperons from the primary ${\pi^-}p \to K^0{\Lambda(1520)}$ channel on $^{12}$C and $^{184}$W target nuclei with
momenta of 0.1--1.45 GeV/c in the laboratory polar angular ranges of 0$^{\circ}$--10$^{\circ}$,
10$^{\circ}$--45$^{\circ}$ and  0$^{\circ}$--45$^{\circ}$ by 1.7 GeV/c $\pi^-$ mesons
as functions of the factor $f$ by which we multiply their collisional width in our model calculations. The lines
are to guide the eye.}
\label{void}
\end{center}
\end{figure}

   Accounting for the HADES spectrometer acceptance, we will consider the $\Lambda(1520)$ momentum
distributions on $^{12}$C and $^{184}$W target nuclei in three laboratory solid angles
${\Delta}{\bf \Omega}_{\Lambda^*}$=$0^{\circ} \le \theta_{\Lambda^*} \le 10^{\circ}$,
$10^{\circ} \le \theta_{\Lambda^*} \le 45^{\circ}$,
$45^{\circ} \le \theta_{\Lambda^*} \le 85^{\circ}$ and
$0 \le \varphi_{\Lambda^*} \le 2{\pi}$. Here, $\varphi_{\Lambda^*}$ is the azimuthal angle of the
$\Lambda(1520)$ momentum ${\bf p}_{\Lambda^*}$ in the laboratory system.
Integrating the full inclusive differential cross section (2) over these ranges,
we can represent the differential cross section for $\Lambda(1520)$ hyperon
production in ${\pi^-}A$ collisions from the direct process (1), corresponding to the
HADES acceptance window
\footnote{$^)$ At HADES the $\Lambda(1520)$ hyperons could be identified via the decays
$\Lambda(1520) \to K^-p$ with a branching ratio of 22.5\%.}$^)$
, in the following form:
\begin{equation}
\frac{d\sigma_{{\pi^-}A\to {\Lambda(1520)}X}^{({\rm prim})}
(p_{\pi^-},p_{\Lambda^*})}{dp_{\Lambda^*}}=
\int\limits_{{\Delta}{\bf \Omega}_{\Lambda^*}}^{}d{\bf \Omega}_{\Lambda^*}
\frac{d\sigma_{{\pi^-}A\to {\Lambda(1520)}X}^{({\rm prim})}
({\bf p}_{\pi^-},{\bf p}_{\Lambda^*})}{d{\bf p}_{\Lambda^*}}p_{\Lambda^*}^2
\end{equation}
$$
=2{\pi}\left(\frac{Z}{A}\right)
\int\limits_{a}^{b}d\cos{{\theta_{\Lambda^*}}}I_{V}[A,\theta_{\Lambda^*}]
\left<\frac{d\sigma_{{\pi^-}p\to K^0{\Lambda(1520)}}(p_{\pi^-},
p_{\Lambda^*},\theta_{\Lambda^*})}{dp_{\Lambda^*}d{\bf \Omega}_{\Lambda^*}}\right>_A,
$$
where $(a,b)=(\cos10^{\circ},1), (\cos45^{\circ},\cos10^{\circ})$ and $(\cos85^{\circ},\cos45^{\circ})$.

   We define now the $\Lambda(1520)$ total in-medium width appearing in Eq. (6) and used in our
calculations of $\Lambda(1520)$ production in $\pi^-A$ reactions. For this width, we employ two
different scenarios [19]: i) no in-medium effects and, correspondingly, the scenario with the free
$\Lambda(1520)$ width (dotted line in Fig. 2); ii) the sum of the free $\Lambda(1520)$ width and its
collisional width $\Gamma_{\rm coll}$ of the type [10, 11, 17] 55($\rho_N/\rho_0)$ MeV, where $\rho_N$ is
the total nucleon density (solid line
\footnote{$^)$ In this scenario the resulting total width of the $\Lambda(1520)$
hyperon reaches the value of about 70 MeV at the normal nuclear matter density $\rho_0$, which
is a factor of $\sim$ 5 larger than the free one.}$^)$
in Fig. 2). In order to study the sensitivity of the $\Lambda(1520)$ hyperon production cross sections
from the channel (1) to its total in-medium width we will also adopt in our calculations 
yet two additional scenarios for this width, in which, as compared to the scenario ii), 
the $\Lambda(1520)$ nominal
collisional width $\Gamma_{\rm coll}$ defined above was artificially multiplied by
factors $f=0.5$ and $f=2$ [17] (dotted-dashed and dashed lines in Fig. 2)
\footnote{$^)$ Evidently, the first two scenarios for the total $\Lambda(1520)$ in-medium width
correspond to the factors $f=0$ and $f=1$.}$^)$
.

 The $\Lambda(1520)$ in-medium width can be extracted from a comparison of the calculated 
(see Eq. (16)) and measured momentum distributions on $^{12}$C and $^{184}$W target nuclei. Additionally,
valuable information concerning the $\Lambda(1520)$ absorption in nuclear matter can be obtained [17, 19]
from a comparison of the calculations with the measured transparency ratio of the $\Lambda(1520)$ hyperon,
normalized to carbon:
\begin{equation}
T_A=\frac{12}{A}\frac{\sigma^{A}_{\Lambda(1520)}}{\sigma^{\rm C}_{\Lambda(1520)}},
\end{equation}
where $\sigma^{A}_{\Lambda(1520)}$ and $\sigma^{\rm C}_{\Lambda(1520)}$ are the inclusive differential (16)
or total
\footnote{$^)$ Obtained by integration of the differential cross sections (16) over $\Lambda(1520)$
momentum.}$^)$
cross sections for $\Lambda(1520)$ production in $\pi^-A$ and $\pi^-$C collisions, respectively.
\begin{figure}[!h]
\begin{center}
\includegraphics[width=18.0cm]{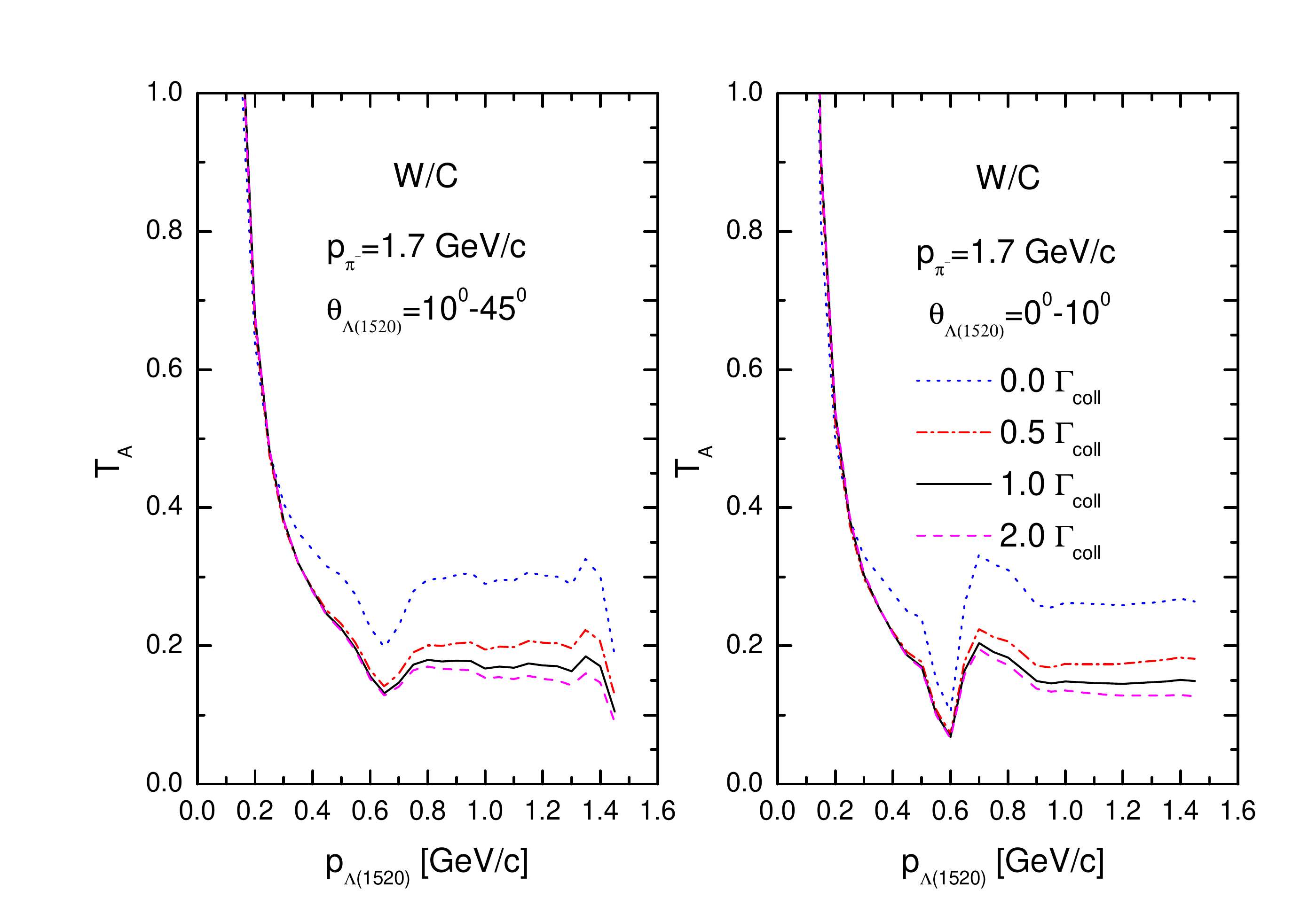}
\vspace*{-2mm} \caption{(color online) Transparency ratio $T_A$ as a function of the $\Lambda(1520)$
momentum for combination $^{184}$W/$^{12}$C as well as for the $\Lambda(1520)$ laboratory polar angular
ranges of 10$^{\circ}$--45$^{\circ}$ (left) and 0$^{\circ}$--10$^{\circ}$ (right), for an incident ${\pi^-}$
meson momentum of 1.7 GeV/c, calculated within different scenarios for the total $\Lambda(1520)$ hyperon
in-medium width where its collisional width is multiplied by the factors indicated in the inset.}
\label{void}
\end{center}
\end{figure}
\begin{figure}[!h]
\begin{center}
\includegraphics[width=18.0cm]{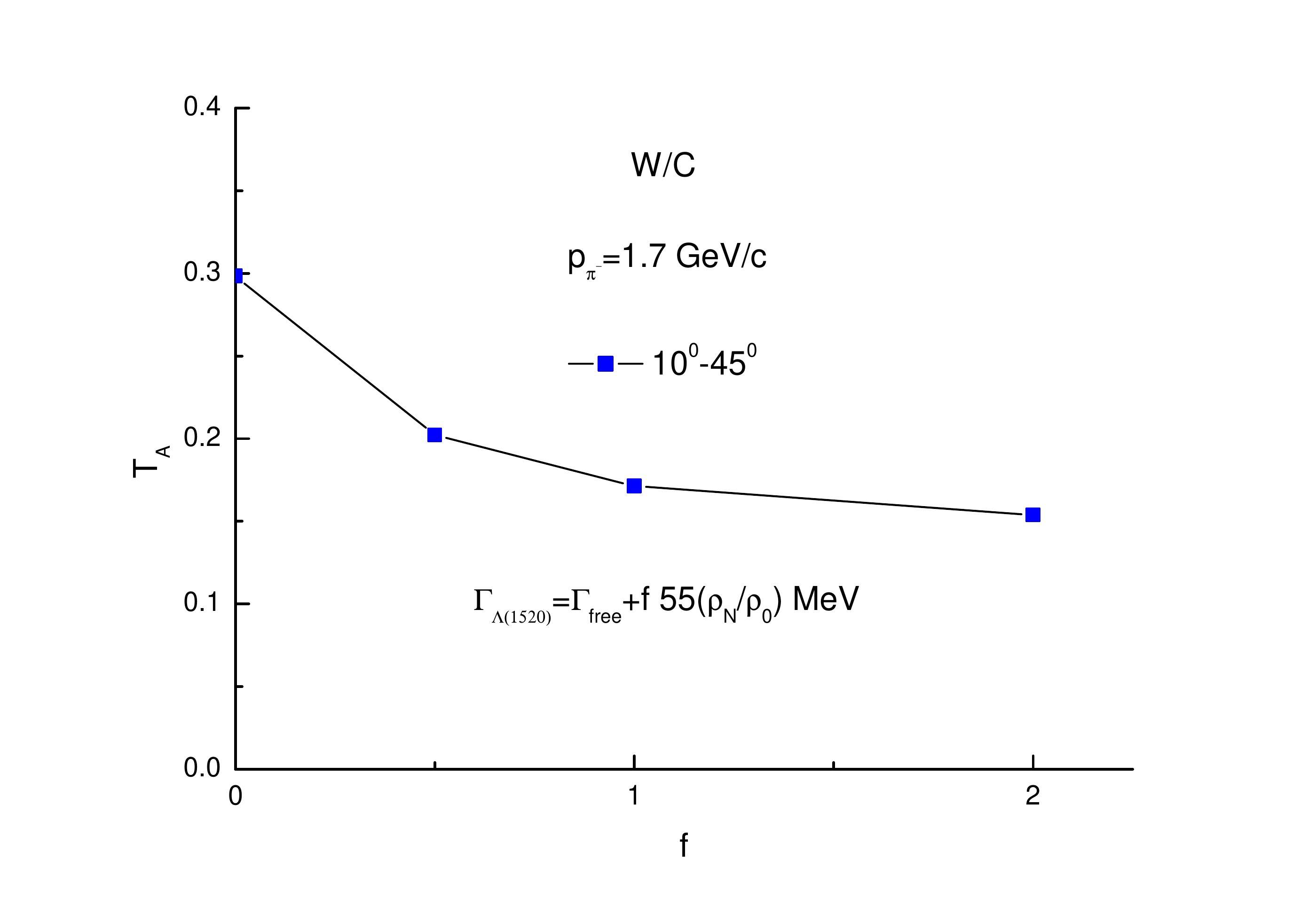}
\vspace*{-2mm} \caption{(color online) Transparency ratio $T_A$ for $\Lambda(1520)$ hyperons from primary
${\pi^-}p \to K^0{\Lambda(1520)}$ reactions at incident ${\pi^-}$ meson momentum of 1.7 GeV/c for 
the target combination
$^{184}$W/$^{12}$C as well as for the $\Lambda(1520)$ laboratory polar angular range of
10$^{\circ}$--45$^{\circ}$ and momentum range of 0.1--1.45 GeV/c as a function of the factor $f$.
The line is to guide the eye.}
\label{void}
\end{center}
\end{figure}

\section*{3 Results and discussion}

\hspace{0.5cm} At first, we consider the absolute $\Lambda(1520)$ momentum differential
cross sections from the direct process (1) in $\pi^-$$^{12}$C and $\pi^-$$^{184}$W collisions for an incident
pion momentum of 1.7 GeV/c. These cross sections were calculated according to Eq. (16) in four
considered scenarios for the total $\Lambda(1520)$ in-medium width (see Fig. 2) at laboratory angles
of 10$^{\circ}$--45$^{\circ}$, 0$^{\circ}$--10$^{\circ}$ and 45$^{\circ}$--85$^{\circ}$. They are given,
respectively, in Figs. 3, 4 and 5. One can see that the $\Lambda(1520)$ hyperons are mainly emitted at
laboratory angles $\le$ 45$^{\circ}$ belonging to the HADES acceptance window. The absolute values of the
differential cross sections have at these angles a well measurable strength $\sim$ 1--100 $\mu$b/(GeV/c)
in the high-momentum region of 0.7--1.4 GeV/c. Here there are a sizeable differences, especially for the
heavy target nucleus $^{184}$W, between the results obtained by using different $\Lambda(1520)$ in-medium
widths under consideration. They are $\sim$ 20--30\% for $^{12}$C and $\sim$ 30--50\% for $^{184}$W between
all calculations corresponding to different choices for this width.

   To see more clearly the sensitivity of the $\Lambda(1520)$ hyperon yield to its in-medium width,
we show in Fig. 6 on a linear scale the ratio between the differential cross sections for $\Lambda(1520)$
hyperon production on $^{12}$C and $^{184}$W nuclei, calculated for different options for its total in-medium
width as presented in Figs. 3, 4, and the respective differential cross sections, determined in the scenario
where the absorption of $\Lambda(1520)$ hyperons in nuclear matter is governed by their free width. It is seen
that there are indeed experimentally distinguishable differences between the considered options for the
$\Lambda(1520)$ in-medium width for both target nuclei and for both forward laboratory polar $\Lambda(1520)$
angular domains. Thus, for example, the $\Lambda(1520)$ momentum distributions are reduced
at collisional width 0.5$\cdot$$\Gamma_{\rm coll}$ by factors of about 1.3 and 2.0 as compared to those obtained without this width at momentum of 1.0 GeV/c and laboratory angular range of 0$^{\circ}$--10$^{\circ}$ for
$^{12}$C and $^{184}$W target nuclei, respectively. When going from 0.5$\cdot$$\Gamma_{\rm coll}$ to
1.0$\cdot$$\Gamma_{\rm coll}$, the corresponding reduction factors are of about 1.2 and 1.4 at these
$\Lambda(1520)$ momentum and angular range; while they are about 1.4 and 1.5
when going from 1.0$\cdot$$\Gamma_{\rm coll}$ to 2.0$\cdot$$\Gamma_{\rm coll}$.

 We, therefore, come to the conclusion that the in-medium properties of $\Lambda(1520)$ hyperons could be
in principle studied at the GSI pion beam facility, using the HADES spectrometer,
through the momentum dependence of their absolute production cross sections in ${\pi^-}A$ interactions
at an initial pion momentum of 1.7 GeV/c.

      The sensitivity of the $\Lambda(1520)$ production differential cross sections at laboratory angles
$\le$ 45$^{\circ}$ in the momentum range $\sim$ 0.7--1.4 GeV/c (where they are largest) to its
in-medium width, shown in Figs. 3, 4, can be also studied from such integral measurements
as the measurements of the total cross sections for $\Lambda(1520)$ production in ${\pi^-}^{12}$C
and  ${\pi^-}^{184}$W reactions by 1.7 GeV/c pions in the full-momentum region of 0.1--1.45 GeV/c.
These cross sections, calculated for the $\Lambda(1520)$ laboratory angular ranges of
0$^{\circ}$--10$^{\circ}$, 10$^{\circ}$--45$^{\circ}$ and 0$^{\circ}$--45$^{\circ}$
by integrating Eq. (16) over the $\Lambda(1520)$ momentum $p_{\Lambda^*}$
in these ranges, are shown in Fig. 7 as functions of the factor $f$ by which the 
$\Lambda(1520)$ collisional width [10, 11, 17] was multiplied in our model calculations. 
One can see that the total cross sections in
the angular regions of 0$^{\circ}$--10$^{\circ}$ and 10$^{\circ}$--45$^{\circ}$ are practically the same
for both target nuclei. They, as well as the sum of them (angular domain of 0$^{\circ}$--45$^{\circ}$)
reveal some sensitivity to the total $\Lambda(1520)$ in-medium width. Thus, the $\Lambda(1520)$
total cross sections in these three angular domains are reduced
at collisional width 0.5$\cdot$$\Gamma_{\rm coll}$ by factors of about 1.3 and 1.9 as compared to
those obtained without this width for $^{12}$C and $^{184}$W target nuclei, respectively.
When going from 0.5$\cdot$$\Gamma_{\rm coll}$ to 1.0$\cdot$$\Gamma_{\rm coll}$,
the corresponding reduction factors are about 1.2 and 1.4; and they are about 1.3 and 1.5
when going from 1.0$\cdot$$\Gamma_{\rm coll}$ to 2.0$\cdot$$\Gamma_{\rm coll}$.
Therefore, a comparison of the "integral" results, presented in Fig. 7, with the respective
experimentally determined total $\Lambda(1520)$ hyperon production cross sections will also allow one
to obtain information about its in-medium width.

   Fig. 8 shows the momentum dependence of the transparency ratio $T_A$ for the target combination W/C
for $\Lambda(1520)$ hyperons produced in the direct reaction channel (1) at laboratory angles of
0$^{\circ}$--10$^{\circ}$ and 10$^{\circ}$--45$^{\circ}$ by 1.7 GeV/c pions. The transparency ratio is
calculated on the basis of Eq. (17) using the results obtained for the adopted options
of the $\Lambda(1520)$ in-medium width and given in Figs. 3, 4. It is seen from this figure that there are
measurable changes $\sim$ 30\% in the quantity $T_A$ only between calculations corresponding to the cases
when the loss of $\Lambda(1520)$ hyperons in nuclear matter is determined by their free width and by the sum
of this width and collisional width of the type 0.5$\cdot$$\Gamma_{\rm coll}$ at the momenta of interest
$\sim$ 0.7--1.4 GeV/c. On the other hand, the differences between the choices
0.5$\cdot$$\Gamma_{\rm coll}$ and 1.0$\cdot$$\Gamma_{\rm coll}$,
1.0$\cdot$$\Gamma_{\rm coll}$ and 2.0$\cdot$$\Gamma_{\rm coll}$ for $\Lambda(1520)$ collisional width are almost
insignificant. They are $\sim$ 10--15\%. This means that the momentum dependence of the transparency ratio
cannot be employed for reliable determination of the $\Lambda(1520)$ in-medium width from
the HADES near-threshold $\Lambda(1520)$ production measurements in ${\pi^-}A$ interactions.
The transparency ratio $T_A$ exhibits dips at momenta $\sim$ 0.6--0.7 GeV/c. This can be explained by the fact that
the differential cross sections for $\Lambda(1520)$ production on $^{184}$W target nucleus "is bent down"
at these momenta (see Figs. 3, 4) due to the off-shell kinematics of the first chance ${\pi^-}p$ collision
and the role played by the nucleus-related effects such as the struck target proton binding and Fermi motion,
encoded in the nuclear spectral function $P_A({\bf p}_t,E)$. The spectral functions for $^{12}$C and
$^{184}$W, adopted in the calculations, are different [21, 23--25].

  In Fig. 9 we show the "integral" transparency ratio $T_A$ for the target combination W/C for $\Lambda(1520)$
hyperons produced at laboratory angles of 10$^{\circ}$--45$^{\circ}$ and momenta of 0.1--1.45 GeV/c
by 1.7 GeV/c pions as a function of the factor $f$ by which their nominal collisional width predicted
in [10, 11, 17] is multiplied in our calculations. This "integral" quantity is calculated according
to Eq. (17) using the results for the total $\Lambda(1520)$ production cross sections presented in Fig. 7.
One can see that the differences between all calculations corresponding to different choices of the
$\Lambda(1520)$ in-medium width are similar to those of Fig. 8.

   Taking into account the above consideration, we can conclude that the $\Lambda(1520)$ absolute
momentum distribution measurements in near-threshold $\pi^-$$^{12}$C and $\pi^-$$^{184}$W reactions
might allow one to shed light on the $\Lambda(1520)$ in-medium width. However, its relative yield
(both "differential" and "integral") in these reactions cannot serve as reliable tool for determining
this width.

\section*{4 Conclusions}

\hspace{0.5cm} In the present paper we study the pion-induced inclusive $\Lambda(1520)$ hyperon
production from $^{12}$C and $^{184}$W target nuclei near threshold
within a nuclear spectral function approach accounting for incoherent primary $\pi^-$ meson--proton
${\pi^-}p \to K^0\Lambda(1520)$ production processes. We calculate the absolute differential and total cross
sections for production of $\Lambda(1520)$ hyperons off these nuclei at laboratory angles of
0$^{\circ}$--10$^{\circ}$, 10$^{\circ}$--45$^{\circ}$ and 45$^{\circ}$--85$^{\circ}$
by 1.7 GeV/c $\pi^-$ mesons as well as their relative (transparency ratio) differential and integral
yields for four scenarios of the $\Lambda(1520)$ total in-medium width. We demonstrate that
these absolute observables, contrary to the relative ones, reveal some sensitivity to
the $\Lambda(1520)$ in-medium width. Therefore, their measurement in a dedicated experiment at
the GSI pion beam facility will allow to shed light on this width.
\\
\\
{\bf Acknowledgments}
\\
\\
The authors would like to thank Volker Metag for careful reading of the manuscript and valuable
comments on it.
\\
\\

\end{document}